# Millimeter-Thick Single-Walled Carbon Nanotube Forests: Hidden Role of Catalyst Support

Suguru Noda[1*], Kei Hasegawa[1], Hisashi Sugime[1], Kazunori Kakehi[1],

Zhengyi Zhang[2], Shigeo Maruyama[2] and Yukio Yamaguchi[1]

[1] Department of Chemical System Engineering, School of Engineering, The University of Tokyo, 7-3-1 Hongo, Bunkyo-ku, Tokyo 113-8656, Japan

[2] Department of Mechanical Engineering, School of Engineering, The University of Tokyo, 7-3-1 Hongo, Bunkyo-ku, Tokyo 113-8656, Japan

A parametric study of so-called "super growth" of single-walled carbon nanotubes (SWNTs) was done by using combinatorial libraries of iron/aluminum oxide catalysts. Millimeter-thick forests of nanotubes grew within 10 min, and those grown by using catalysts with a thin Fe layer (about 0.5 nm) were SWNTs. Although nanotube forests grew under a wide range of reaction conditions such as gas composition and temperature, the window for SWNT was narrow. Fe catalysts rapidly grew nanotubes only when supported on aluminum oxide. Aluminum oxide, which is a well-known catalyst in hydrocarbon reforming, plays an essential role in enhancing the nanotube growth rates.

---

[*] Corresponding author. E-mail address: noda@chemsys.t.u-tokyo.ac.jp

Soon after the realizations of the vertically-aligned single-walled carbon nanotube (VA-SWNT) forests[1] by alcohol chemical vapor deposition (ACCVD),[2] many groups achieved this morphology of nanotubes by several tricks in CVD conditions.[3-6] Among these methods, the water-assisted method, the so-called "super growth" method,[3] realized an outstanding growth rate of a few micrometers per second, thus yielding millimeter-thick VA-SWNT forests. Despite its significant impact on the nanotube community, no other research groups have been successful in reproducing "super growth". Later, the control of the nominal thickness of Fe in the Fe/ $Al_2O_3$ catalyst was shown crucial for controlling the number of walls and diameters of the nanotubes.[7] In this work, we carried out a parametric study of this growth method by using a combinatorial method that we previously developed for catalyst optimization.[8,9]

Si wafers that had a 50-nm-thick thermal oxide layer and quartz glass substrates were used as substrates, and Fe/ $SiO_2$, Fe/$Al_2O_x$, and Fe/$Al_2O_3$ catalysts were prepared by sputter deposition on them. An $Al_2O_x$ layer was formed by depositing 15-nm-thick Al on the substrates, and then exposing the layer to air. A 20-nm-thick $Al_2O_3$ layer was formed by sputtering an $Al_2O_3$ target. Then, Fe was deposited on $SiO_2$, on $Al_2O_x$, and on $Al_2O_3$. In some experiments, gradient-thickness profiles were formed for Fe by using the combinatorial method previously described.[9] The catalysts were set in a tubular, hot-wall CVD reactor (22-mm inner diameter and 300-mm length), heated to a target temperature (typically 1093 K), and kept at that temperature for 10 min while being exposed to 27 kPa $H_2$/ 75 kPa Ar at a flow rate of 500 sccm, to which $H_2O$ vapor was added at the same partial pressure as for the CVD condition (i.e., 0 to 0.03 kPa). During this heat treatment, Fe formed into a nanoparticle structure with a diameter and areal density that depended on the initial Fe thickness.[8] After the heat treatment, CVD was carried out by switching the $H_2$/ $H_2O$ /Ar gas to $C_2H_4$/ $H_2$/ $H_2O$/ Ar. The standard condition was 8.0 kPa $C_2H_4$/ 27 kPa $H_2$/ 0.010 kPa $H_2O$/ 67 kPa Ar and 1093 K. The samples were analyzed by using transmission electron microscopy (TEM) (JEOL JEM-2000EX) and micro-Raman scattering spectroscopy (Seki Technotron, STR-250) with an excitation wavelength at 488 nm.

Figure 1a shows a photograph of the nanotubes grown for 30 min under the standard condition. Nanotubes formed forests that were about 2.5 mm thick. The taller nanotubes at the edge compared with those at the center of the substrates indicate that the nanotube growth rate was limited by the diffusion of the growth species through the millimeter-thick forests of nanotubes. Figure 1b shows a TEM image of the as-grown sample shown in the center of Fig. 1a. The nanotubes were mostly SWNTs. These figures show that "super growth" was achieved. Although catalysts with thicker Fe layer ($\geq$ 1 nm) yielded rapid growth for a wide range of CVD conditions, mainly multi-walled nanotubes (MWNTs) formed instead of SWNTs. Rapid growth of SWNTs requires complicated optimization of the CVD conditions, i.e., $C_2H_4$/ $H_2$/ $H_2O$ pressures and the growth temperature, because the thinner layer of Fe catalysts (around 0.5 nm) yielded rapid SWNT growth under a narrow window near the standard condition.

The effect of the catalyst supports on the nanotube growth was also studied here. Figure 2a shows normal photographs of nanotubes grown by using three types of combinatorial catalyst libraries; i.e., Fe/ $SiO_2$, Fe/ $Al_2O_x$, and Fe/ $Al_2O_3$. For the Fe/ $SiO_2$ catalyst, the surface was slightly darker at regions with 0.4- to 0.5-nm-thick Fe. For the Fe/ $Al_2O_x$, and Fe/ $Al_2O_3$ catalyst, the result was completely different; nanotube forests even thicker than the substrates were formed within 10 min. Differences also were evident between the catalysts with $Al_2O_x$ and $Al_2O_3$ supports. When Fe was relatively thick (> 0.6 nm), nanotube forests grew thick by using either of these two catalysts. When Fe was thinner (≤ 0.6 nm), however, nanotube forests grew thick only by using the Fe/ Al$_2$O$_x$ catalyst (Fig. 2b). Figure 2c shows Raman spectra taken at several locations for each catalyst library. For Fe/ SiO$_2$, a Raman signal of nanotubes was obtained only when the Fe layer was thin (i.e. ≤ 0.8 nm). The sharp and branched G-band with small D-band and the peaks of radial breathing mode (RBM) indicate the existence of SWNTs. The G/D peak area ratios exceeding 10 indicate that the SWNTs were of relatively good quality. For Fe/ Al$_2$O$_x$, the Raman signal of nanotubes was observed also for a thick Fe region (i.e. ≥ 1.0 nm) with G/D ratios somewhat smaller than the G/D ratios for Fe/ SiO$_2$. The G/D ratio of 10 for the nanotubes by 0.5 nm Fe/ Al$_2$O$_x$ shows that the SWNTs still were of relatively good quality compared with the original "super growth".[3] As the Fe thickness was increased, G/D ratios became smaller because MWNTs became the main product at the thicker Fe regions. For Fe/ Al$_2$O$_3$, the results were similar to those for Fe/ Al$_2$O$_x$ except when the Fe layer was thin (around 0.5 nm) where nanotube forests did not grow. Similar phenomenon was observed also for Co and Ni catalysts; they yielded nanotube forests when supported on an aluminum oxide layer. These results show that an aluminum oxide layer is essential for "super growth", that the growth rate enhancement by Al$_2$O$_x$ might accompany some decrease in the G/D ratio, and that the catalyst Fe layer needs to be thin (< 1 nm for the CVD condition studied here) to grow SWNTs. An Al$_2$O$_x$ catalyst support was more suitable than Al$_2$O$_3$ to grow SWNTs, and the underlying growth mechanism is now under investigation.

    The effect of the H$_2$O vapor on the nanotube growth was studied next. Figure 3a shows the thickness profiles of nanotube forests grown on the Fe/ Al$_2$O$_x$ catalyst

library. In the absence of $H_2O$ vapor, nanotubes grew at the thin Fe region (0.3- to 1-nm thick). Addition of 0.010 kPa $H_2O$, which corresponds to 100 ppmv in the reactant gases, enhanced the nanotube growth, especially at the thicker Fe region (> 0.7 nm). Further addition of $H_2O$ (0.030 kPa), however, inhibited the nanotube growth at the thinner Fe region (0.3- 0.6 nm) where SWNTs grew at lower $H_2O$ partial pressures. Figure 3b shows Raman spectra of these samples. Slight addition of $H_2O$ (0.01 kPa) did not affect the G/D ratio at the thin Fe region (0.5 nm) but decreased the G/D ratio at the thicker region (0.8 and 1.0 nm). Further addition of $H_2O$ (0.03 kPa) significantly decreased the G/D ratio at the whole region of Fe thickness. These results show that the $H_2O$ addition up to a certain level can enhance the nanotube growth rate, but too much addition degrades the nanotube quality.

Considering that alumina and its related materials catalyze hydrocarbon reforming,[10] a possible mechanism for "super growth" is proposed as follows: $C_2H_4$ or its derivatives adsorb onto aluminum oxide surfaces, diffuse on the surface to be incorporated into Fe nanoparticles, and segregate as nanotubes from Fe nanoparticles. $H_2O$ vapor keeps aluminum oxide surface reactive by removing the carbon byproducts, while simultaneously, $H_2O$ reacts with the nanotubes and degrades the quality of the nanotubes. The $C_2H_4/H_2O$ pressure ratio needs to be kept large (790 for the standard condition in this work) as previously reported in ref. 11. The complicated optimization among $C_2H_4$, $H_2$, and $H_2O$ to achieve "super growth" of SWNTs indicates that balancing the carbon fluxes of adsorption onto aluminum oxides, the surface diffusion from aluminum oxides to Fe nanoparticles, and the segregation as nanotubes from Fe nanoparticles is essential to sustain the rapid nanotube growth at a few micrometers per second. During nanotube growth, because the surface of catalyst nanoparticles is mostly covered by nanotubes, nanotube growth can be enhanced by introducing a carbon source not only through the limited open sites on catalyst nanoparticles but also through the catalyst supports whose surface remains uncovered with growing nanotubes. This concept might provide a new route for further development of supported catalysts for nanotube growth.

**Acknowledgements:**

This work is financially supported in part by the Grant-in-Aid for Young Scientists (A), 18686062, 2006, from the Ministry of Education, Culture, Sports, Science and Technology (MEXT), Japan.

**References:**

1) Y. Murakami, S. Chiashi, Y. Miyauchi, M. Hu, M. Ogura, T. Okubo and S. Maruyama: Chem. Phys. Lett. **385** (2004) 298.

2) S. Maruyama, R. Kojima, Y. Miyauchi, S. Chiashi and M. Kohno: Chem. Phys. Lett. **360** (2002) 229.

3) K. Hata, D.N. Futaba, K. Mizuno, T. Nanami, M. Yumura and S. Iijima: Science **306** (2004) 1362.

4) G. Zhong, T. Iwasaki, K. Honda, Y. Furukawa, I. Ohdomari and H. Kawarada: Jpn. J. Appl. Phys. **44** (2004) 1558.

5) L. Zhang, Y. Tan and D.E. Resasco: Chem. Phys. Lett. **422** (2006) 198.

6) G. Zhang, D. Mann, L. Zhang, A. Javey, Y. Li, E. Yenilmez, Q. Wang, J. P. McVittie, Y. Nishi, J. Gibbons and H Dai, Proc. Nat. Acad. Sci. **102** (2005) 16141.

7) T. Yamada, T. Nanami, K. Hata, D.N. Futaba, K. Mizuno, J. Fan, M. Yudasaka, M. Yumura and S. Iijima: Nat. Nanotechnol. **1** (2006) 131.

8) S. Noda, Y. Tsuji, Y. Murakami and S. Maruyama: Appl. Phys. Lett. **86** (2005) 173106.

9) S. Noda, H. Sugime, T. Osawa, Y. Tsuji, S. Chiashi, Y. Murakami and S. Maruyama: Carbon **44**, (2006) 1414.

10) S.E. Tung and E, Mcininch: J. Catal. **4** (1965) 586.

11) D.N. Futaba, K. Hata, T. Yamada, K. Mizuno, M. Yumura and S. Iijima: Phys. Rev. Lett. 95 (2005) 056104.

**Figure Captions:**

Fig. 1. Typical nanotubes grown in this work. (a) Normal photographs of nanotube forests grown on Fe/ Al$_2$O$_x$ for 30 min under the standard condition (8.0 kPa C$_2$H$_4$/ 27 kPa H$_2$/ 0.010 kPa H$_2$O/ 67 kPa Ar and 1093 K). Fe catalyst thickness was uniform at 0.45 nm (left sample), 0.50 nm (middle), and 0.55 nm (right). (b) TEM image of nanotubes in Fig. 1a grown using 0.50-nm-thick Fe catalysts. Insets show the enlarged images (2.5x) of nanotubes.

Fig. 2. Effect of support materials for Fe catalyst on nanotube growth. Nanotubes were grown for 10 min under the standard condition. (a) Photographs of nanotubes grown by using combinatorial catalyst libraries, which had a nominal Fe thickness profile ranging from 0.2 nm (at left on each sample) to 3 nm (right) formed on either SiO$_2$, Al$_2$O$_x$, or Al$_2$O$_3$. (b) Relationship between the thickness of nanotube forest (shown in Fig. 2a) and the nominal Fe thickness of the catalyst. (c) Raman spectra of the same samples. Intensity at the low wavenumber region (< 300 cm$^{-1}$) is shown magnified by a factor of 5x in this figure. Declined background signals in some of the RBM spectra (e.g., 0.5, 0.8-nm-Fe/ SiO$_2$ and 0.5-nm-Fe/ Al$_2$O$_3$) were due to the signal from SiO$_2$ substrates passing through the thin nanotube layer.

Fig. 3. Effect of H$_2$O vapor on the nanotube growth. Nanotubes were grown using Fe/ Al$_2$O$_x$ combinatorial catalyst libraries for 10 min under the standard condition except for H$_2$O partial pressures. (a) Relationship between the thickness of nanotube forest and the nominal Fe thickness of the catalyst at different H$_2$O partial pressures. (b) Raman spectra of the same samples. Intensity at the low wavenumber region (< 300 cm$^{-1}$) is shown magnified by a factor of 5x in this figure.

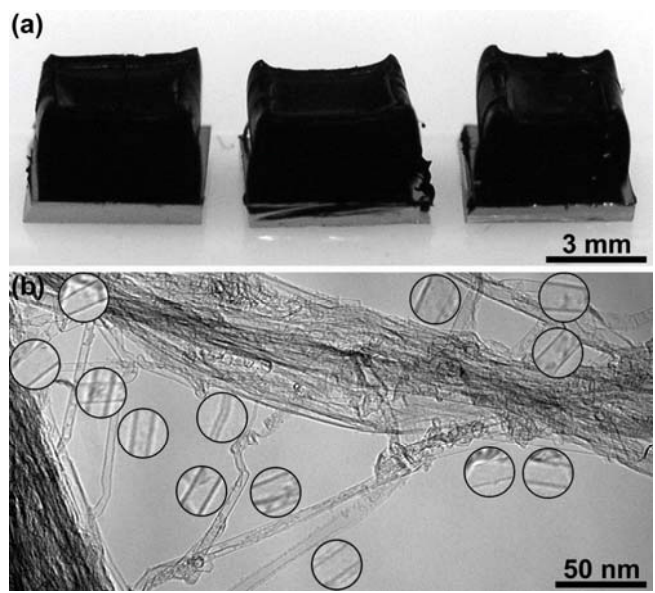

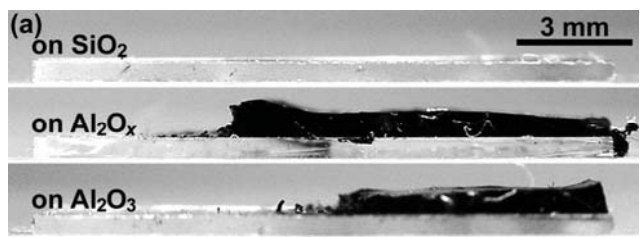
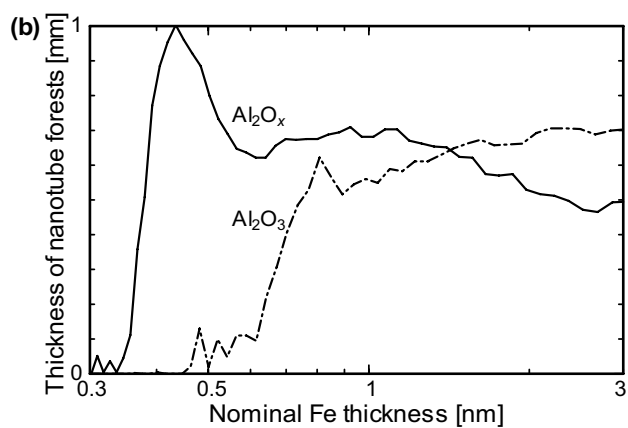
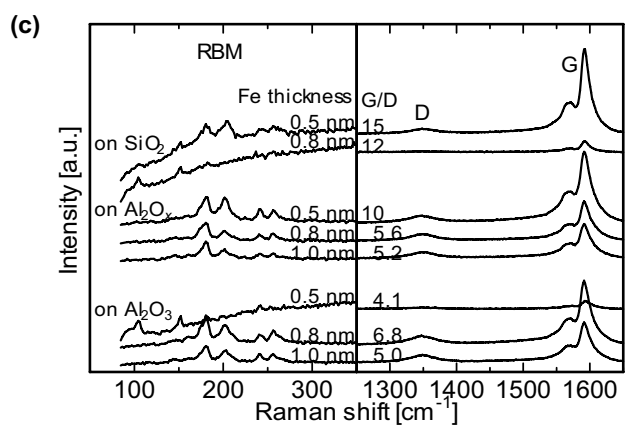

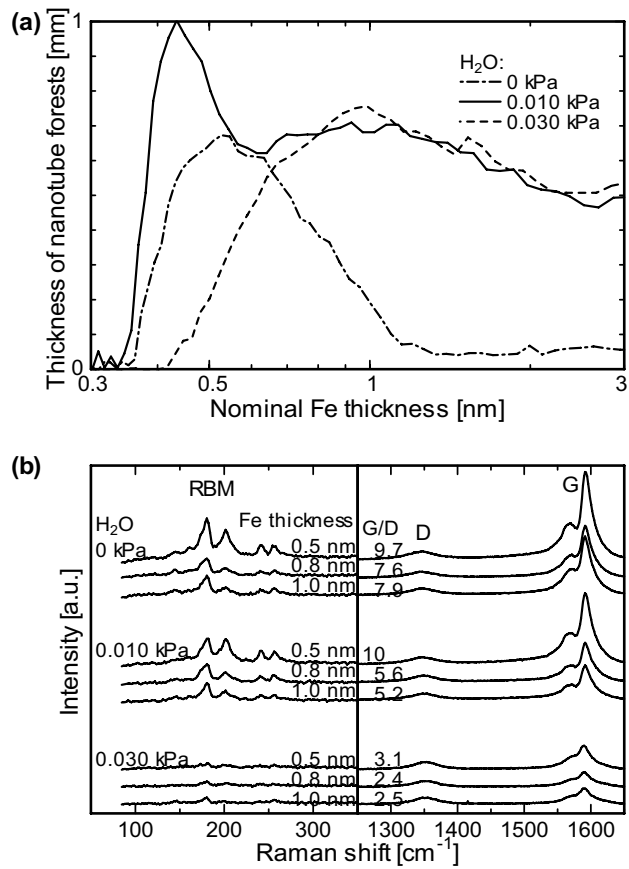